
%
%

\input harvmac
\voffset=0.5truein
\overfullrule=0pt

\def\centeron#1#2{{\setbox0=\hbox{#1}\setbox1=\hbox{#2}\ifdim
	\wd1>\wd0\kern.5\wd1\kern-.5\wd0\fi
	\copy0\kern-.5\wd0\kern-.5\wd1\copy1\ifdim\wd0>\wd1
	\kern.5\wd0\kern-.5\wd1\fi}
}

\def\frac#1#2{{\textstyle{#1\over #2}}}
\def\Sl#1{\centeron{$#1$}{$/$}}
\def\hf{\frac12}
\def\vph{{\vec\p}}
\def\vt{{\vec\vartheta}}

\def\zb{\bar{z}}
\def\over{\above0.15pt}
\def\tr{{\rm tr}}
\def\Tr{{\rm Tr}}

\def\log{{\rm log}}
\def\sign{{\rm sign}}
\def\Pit{{\tilde\Pi}}
\def\omv{{\vec\omega}}
\def\omh{{\hat\omega}}

\let\lm=\xi
\let\s=\sigma
\let\p=\varphi
\let\e=\varepsilon
\let\la=\lambda
\let\La=\Lambda
\let\d=\partial
\let\a=\alpha
\let\b=\beta
\let\g=\gamma
\let\l=\left
\let\r=\right
\let\dg=\dagger


\font\blackboard=cmbx10
\def\ZZ{{\hbox{\blackboard Z}}}


\newif\ifdraft \draftfalse
\newcount\eqnum
\eqnum=0
\def\eq{\global\advance\eqnum by1 \eqno(\secsym\the\eqnum)}
\def\eqlabel#1{\global\advance\eqnum by1
	\ifdraft\eqno{{\tt [\string#1]}(\secsym\the\eqnum)}\else
	\eqno{(\secsym\the\eqnum)}\fi
	\xdef#1{\secsym\the\eqnum}}


\def\pl#1#2#3{{Phys. Lett.} {\bf #1} (#2) #3}
\def\np#1#2#3{{Nucl. Phys.} {\bf #1} (#2) #3}
\def\pr#1#2#3{{Phys. Rev.} {\bf #1} (#2) #3}
\def\prl#1#2#3{{Phys. Rev. Lett.} {\bf #1} (#2) #3}
\def\ijmp#1#2#3{{Int. J. Mod. Phys.} {\bf #1} (#2) #3}

\lref\WEINBERG{S. Weinberg, {\it Prog. Theor. Phys. Suppl.} {\bf 86} (1986)
43.} \lref\DJT{S. Deser, R. Jackiw and S. Templeton, {\it Ann. Phys.} {\bf 140}
(1982) 372.}
\lref\HST{Z. Hlousek, D. S\'en\'echal and S.-H. H. Tye, \pr{D41}{1990}{3773}.}
\lref\DR{S. Deser and A.N. Redlich, \prl{61}{1988}{1541}.}
\lref\DM{N. Dorey and N.E. Mavromatos, \pl{B250}{1990}{107};
preprint CERN-TH.6278/91.}
\lref\CW{Y.-H. Chen and F. Wilczek, \ijmp{B3}{1989}{1252}.}
\lref\CCW{M. Carena, S. Chaudhuri and C.E.M. Wagner, \pr{D42}{1990}{2120}.}
\lref\BDP{K.S. Babu, A. Das and P. Panigrahi, \pr{D36}{1987}{3725}.}
\lref\KR{A. Kovner and B. Rosenstein, \pr{B42}{1990}{4748}.}
\lref\SW{G.W. Semenoff and N. Weiss, \pl{250}{1990}{117}.}
\lref\CHN{S. Chakravarty, B. Halperin and D. Nelson, \pr{B39}{1989}{2344}.}
\lref\COLEMAN{S. Coleman, {\it Aspects of Symmetry}, Cambridge University
Press (Cambridge) 1985.}
\lref\CH{S. Coleman and B. Hill, \pl{B159}{1985}{184}.}
\lref\BL{T. Banks and J.D. Lykken, \np{B336}{1990}{500}.}


\Title{}{\vbox {
\centerline{Chern-Simons Superconductivity without Parity Violation}
}}
\centerline{D. S\'en\'echal}
\bigskip\centerline{\it D\'epartement de Physique,}
\centerline{\it Universit\'e de Sherbrooke,}
\centerline{\it Sherbrooke (Qu\'ebec) G1K 7P4  Canada.}
\vskip .2in

\noindent
\medskip
\noindent
We argue that a simple Yukawa coupling between the $O(3)$ nonlinear
$\s$-model and charged Dirac fermions leads, after one-loop quantum
corrections, to a Meissner effect, in the disordered phase of the
nonlinear $\s$-model.

\Date{(July 1992, revised Sept. 1992)}
\baselineskip=12pt
\newsec{Introduction}

Superconductivity is characterized by a spontaneous breakdown of
electromagnetic gauge invariance.
Electromagnetism is then effectively described by the London Lagrangian:
$$ {\cal L}_L~=~-\frac14 F_{\mu\nu}F^{\mu\nu} +
\hf (\d_\mu\phi-mA_\mu)(\d^\mu\phi-mA^\mu) \eqlabel\London $$
where $\phi$ is a scalar field transforming like $\phi\to\phi+m\La$
under the gauge transformation $A_\mu\to A_\mu+\d_\mu\La$.
Such an action, in which the electromagnetic field $A_\mu$ acquires a mass
$m$, is usually obtained through the Higgs mechanism.

However, in $(2+1)$ dimensions this mechanism may manifest itself in a
curious way.
It is known \DJT\ that a gauge-invariant electromagnetic mass may be
introduced explicitly in the Lagrangian, via a Chern-Simons term.
Indeed, the Lagrangian density
$$ {\cal L}~=~-\frac14 F_{\mu\nu}F^{\mu\nu} +
\frac12 m\e^{\mu\nu\la}A_\mu\d_\nu A_\la\eqlabel\CSa $$
describes photons of mass $m$. It has, however, the unattractive feature
of violating parity (P) and time-reversal (T).
A solution to this problem is the introduction (to be justified, of course)
of a second Abelian gauge field $a_\mu$, coupled to the electromagnetic
field via a {\it mixed} Chern-Simons term:
$$ {\cal L}~=~-\frac14 F_{\mu\nu}F^{\mu\nu} -\frac14 f_{\mu\nu}f^{\mu\nu}
-m\e^{\mu\nu\la}A_\mu\d_\nu a_\la \eqlabel\CSb $$
where $f_{\mu\nu}=\d_\mu a_\nu - \d_\nu a_\mu$ and where the two gauge
fields $a_\mu$ and $A_\mu$ have opposite intrinsic parities.
P and T are manifestly conserved, and it is
straightforward to show that the excitations of this theory have mass $m$.
Indeed, in terms of the two fields $A^\pm_\mu=(A_\mu\pm a_\mu)/\sqrt2$ the
Lagrangian (\CSb) is simply a sum of two expressions of the type (\CSa) for
$A^+_\mu$ and $A^-_\mu$ respectively, with opposite masses, and wherein
$A^+_\mu$ and $A^-_\mu$ are exchanged under parity.
In other words, the photon mixes with the auxiliary gauge field $a_\mu$ and
acquires a mass in the process.
It has been shown in \DM\ that the Lagrangian (\CSb) is functionally
equivalent to the London form (\London).
A similar calculation has been presented in \BL. The reader willing to
perform this straightforward demonstration will find hints in these
works.

The aim of this paper is to argue that a Lagrangian of the form (\CSb) arises
in the 1-loop effective action of the $O(3)$ nonlinear $\s$-model coupled to
charged Dirac fermions via a simple Yukawa coupling.
This result is of potential interest in the theory of high-temperature
superconductivity, although our treatment will be purely relativistic.

This paper is organized as follows. In section 2 we briefly describe the
$O(3)$ nonlinear $\s$-model.
We explain the origin of the auxiliary gauge field $a_\mu$ and show how this
field acquires a Maxwell-like dynamics ($f_{\mu\nu}f^{\mu\nu}$)  at one loop
within the nonlinear $\s$-model.
In section 3 we couple the model to charged Dirac fermions and show how
fermion loops give rise to a mixed Chern-Simons term, in addition to
renormalizing the two Maxwell terms.
Discussions similar in spirit to this one have already appeared in the
literature \KR\SW, especially in Ref. \DM. In this work, however, we stay
within a continuum theory, starting from a simple continuum model of
interactions with fermions. Our conclusions are similar, though not
entirely identical, with those of Ref. \DM.

\newsec{The $O(3)$ nonlinear $\s$-model}

\subsec{Generalities}

The defining field of the $O(3)$ nonlinear $\s$-model is a unit vector
$\vph$ living on the 2-sphere $S_2$.
The Lorentz invariant Lagrangian of the model is
$$ {\cal L}_\s ~=~ {1\over 2\a^2} \d_\mu\vph\cdot\d^\mu\vph \eqlabel\NLSa $$
Despite its quadratic appearance this Lagrangian does not describe a free
theory, because of the constraint $\vph^2=1$: the field $\vph$ has really
two degrees of freedom.
In condensed matter applications this field represents the staggered
magnetization of antiferromagnets, and for this reason we will refer to it
as {\it spin}.
The Lagrangian (\NLSa) is symmetric under global $O(3)$ rotations of the spin
$\vph$.
At tree-level this internal symmetry is spontaneously broken to $U(1)$ by a
ground state in which all spins are uniformly aligned.
Consequently, the lowest excitations are massless Goldstone bosons (spin
waves) of two different polarizations, propagating at unit speed.

In addition to spin waves, the $O(3)$ nonlinear $\s$-model admits topological
solitons.
They originate from the homotopy $\pi_2(S_2)=\ZZ$. Indeed, a
configuration of $\vph$ with finite energy must have all the spins aligned at
infinity, in all directions. In practice all the points at infinity can be
identified, thereby compactifying the plane into a 2-sphere. A finite energy
configuration is then a map $\vph:S_2\to S_2$, and these maps fall into
different homotopy classes, indexed by a winding number $Q$ (the topological
charge).
The minimum energy configurations in the homotopy class of charge
$Q\not=0$ are solitons.
The solitonic configuration centered at the origin and aligned with $\hat{z}$
at infinity is ($N=|Q|$)
$$ \vph = \l({2r^N\over r^{2N}+1}\cos(Q\phi)~,~{2r^N\over
r^{2N}+1}\sin(Q\phi)~,
in cylindrical coordinates $(r,\phi)$.
Its mass is $4\pi |Q|/\a^2$.

To the soliton charge corresponds a topological current:
$$
j^\mu ~=~ {1\over 8\pi}\e^{abc}\e^{\mu\nu\la}\p^a\d_\nu\p^b\d_\la\p^c
\eqlabel\topoC $$
conserved without using the equations of motion.
For this reason $j^\mu$ may be expressed as
$$ j^\mu = \e^{\mu\nu\la} \d_\nu a_\la \eqlabel\auxGF $$
where $a_\mu$ is an auxiliary gauge field, fixed by $j^\mu$ (hence by $\vph$)
up to a gauge transformation.
It is this field, after some rescaling, that will appear in (\CSb).
The kinetic term $f_{\mu\nu}f^{\mu\nu}$ is then proportional to the
current-current term $j^\mu j_\mu$, while the mixed Chern-Simons term
is nothing but a coupling $j^\mu A_\mu$ between the soliton current
and the electromagnetic field, meaning that the soliton has acquired
an electric charge.
These effects are revealed by one-loop quantum corrections.

The constant $\a$ has the meaning of a coupling,
since the theory becomes free as $\a\to 0$.
Indeed, the rescaling $\vt=\a\vph$ gives the kinetic term for $\vt$ a canonical
normalization and the nonlinear constraint becomes  $\vt^2=\a^{-2}$.
As $\a\to0$, the sphere on which $\vt$ lives is sent to infinity
and looks, around a given point $\vt$, like a plane.
Fields living on that plane are free, i.e., the constraint is linearized.
In this limit ($\a\to 0$) the solitons become infinitely massive and decouple
from the system.

Finally, let us remark that the $O(3)$ nonlinear $\s$-model is not a
renormalizable field theory in $(2+1)$ dimensions, although it is in
$(1+1)$ dimensions.
This needs not disturb us, since we will be interested in long-distance
behavior, unaffected by the plethora of multi-derivative terms brought
about by quantum corrections.

\subsec{The $CP^1$ Formulation}

In order to show how quantum corrections confer a Maxwell dynamics to
the auxiliary gauge field $a_\mu$ introduced in (\auxGF) we translate
the nonlinear $\s$-model into the $CP^1$ model.
A two-component complex boson $z\equiv(z_1,z_2)$ is introduced, in terms
of which $\vph$ is expressed as
$$ \vph ~=~ \zb \vec{\s}z\eq $$
The correspondence is not one-to-one since any local phase change
$z\to e^{i\theta}z$ leaves $\vph$ unchanged.
The constraint $\vph^2=1$ translates into $\zb z=1$, while
the Lagrangian (\NLSa) may be written as
$$ {\cal L}_\s ~=~ {2\over\a^2}(\d_\mu-ia_\mu)\zb
(\d^\mu+ia^\mu)z\eqlabel\CPone $$
wherein the auxiliary vector field $a_\mu$ is to be varied independently.
It is straightforward to show that the equation of motion $a_\mu=\zb\d_\mu z$
is equivalent to (\topoC) and (\auxGF).
The Lagrangian (\CPone) has a local $U(1)$ symmetry, $a_\mu$ playing the r\^ole
of a gauge field (without Maxwell term at the classical level).
This extra symmetry comes from the phase arbitrariness of the boson doublet
$z$.

In the quantum theory, the constraint $\zb z=1$ is implemented by a functional
delta function $\delta(\zb z-1)$ within the integration measure, which may
then be converted into an integral over a Lagrange multiplier $\lm$: The
partition function of the $CP^1$ model is then
$$ Z~=~\int[da_\mu][d\zb dz][d\lm]~\exp i{2\over\a^2}\bigg\{-\lm (\zb z-1)
+(\d_\mu-ia_\mu)\zb(\d^\mu+ia^\mu)z \bigg\} \eq $$
In $(1+1)$ dimensions, it is well-known \COLEMAN\ that integrating the
$z$ field produces an effective potential $V(\lm)$ which admits a
minimum at some value $\lm_0$, and that accordingly the $z$-quanta, which
are nothing but spin-waves, become massive. Therefore they
are no longer Goldstone bosons: the global $su(2)$ symmetry is no longer
spontaneously broken, but has been dynamically restored. Integrating the
$z$ field also produces a Maxwell term (with the right sign) for the
auxiliary gauge field $a_\mu$.

Deser and Redlich \DR\ have extended
Coleman's original calculation to $(2+1)$ dimensions. In that case the
effective potential (for $\lm>0$) turns out to be
$$ V(\lm) = {\lm\over 2\a^2}- {M\lm\over 8\pi} + {\lm^{3/2}\over 12\pi}\eq$$
where $M$ is the subtraction point and $\a$ is now the renormalized coupling.
If $\a>\sqrt{4\pi/M}$ this potential admits a minimum at
$\lm_0=(M-4\pi/\a^2)^2$, and $\sqrt{\lm_0}$ is the effective mass of the $z$
quanta. As in $(1+1)$ dimensions, the symmetry has been dynamically restored:
we are in a disordered (or symmetric) phase.
If, on the other hand, $\a<\sqrt{4\pi/M}$, the effective potential does
not admit a minimum for positive $\lm$. We are then in a phase of
broken symmetry in which the spin-waves are massless.
This interpretation is in agreement with the renormalization group
calculations of Chakravarty et al.\CHN.
The calculation of the induced Maxwell term in the symmetric phase is
straightforward.
At one loop we need not take into account the dynamics of the $\lm$ field:
only its vacuum value $\sqrt{\lm_0}$ is important, as a $z$-mass.
Gauge invariance saves us from UV divergences and the vacuum polarization
tensor is
$$\eqalign{
\Pi^{\mu\nu} &= -i\Pi(p^2)\l( g^{\mu\nu}p^2-p^\mu p^\nu\r)\cr
\Pi(0) &= {1\over 12\pi\sqrt{\lm_0}}\cr} \eqlabel\polA $$
Consequently the induced Maxwell term for $a_\mu$ is
$${\cal L}_M=-{1\over 48\pi\sqrt{\lm_0}}f_{\mu\nu}f^{\mu\nu}\eqlabel\kinA$$
Note that this kinetic term has the correct sign only in the symmetric
phase ($\sqrt{\lm_0}>0$).
This result is of course independent of the $CP^1$ formalism used to derive it:
As mentioned above, the Maxwell term is nothing but a current-current term
$j_\mu j^\mu$.
It could have been obtained, albeit with more effort, within the original
$\s$-model formulation.

\newsec{Fermions}

Let us couple $\vph$ to an $su(2)$ doublet of charged, massless Dirac
fermions via a Yukawa coupling:
$$ {\cal L}_f = \bar\psi(i\Sl\d-g\vph\cdot\vec\s-e_0\Sl{A})\psi
\eqlabel\FerA $$
$e_0$ is the fermion's bare electric charge, and has the dimensions
of (mass)$^{1/2}$.
In $(2+1)$ dimensions a single Dirac fermion has 2 components. A
representation of the Dirac matrices is given by $\g^0=\tau_3$, $\g^1=i\tau_1$
and $\g^2=i\tau_2$, where the $\tau_a$ are Pauli matrices, but acting on a
different space from $\s_a$.
The internal $su(2)$ symmetry is incorporated by
considering a doublet of Dirac spinors (i.e. 4 components in all). The fermion
$\psi$ thus carries a Dirac index and a `spin' (or $su(2)$) index. The Yukawa
coupling is the simplest possible interaction to introduce, with the least
number of derivatives.

We will now argue that fermion-induced one-loop quantum corrections
lead to a mixed Chern-Simons term like in (\CSb).
In order to perform this calculation we will again use a change of variables
from the original $\s$-model.

\subsec{The Spin Rotation}

Let $U$ be a $su(2)$ matrix that rotates the matrix
$\p=\vph\cdot\vec\s$ towards $\s_3$, and let $\chi$ be the fermion
obtained by applying the same rotation to $\psi$:
$$ \p = U^{-1}\s_3 U \qquad\qquad \chi=U\psi \eqlabel\rotU $$
In terms of the rotated variables, the fermion Lagrangian becomes
$$ {\cal L}_f = \bar\chi(i\Sl\d -g\s_3 - \Sl{B} -e_0\Sl{A})\chi
\eqlabel\FerB $$
where we have introduced an auxiliary, curvature-free gauge field:
$$\eqalign{
B_\mu &= -iU\d_\mu U^{-1}\cr
&= 2\pi a_\mu\s_3 + b_\mu\s_- + b^*_\mu\s_+ \qquad
\s_\pm=\hf(\s_1\pm i\s_2)\cr}\eq $$
The three internal components $a_\mu$, $b_\mu$ and $b^*_\mu$ of $B_\mu$ are
related to each other by the curl-free condition
$$ \d_\mu B_\nu-\d_\nu B_\mu +i[B_\mu,B_\nu]=0\eq $$
The $\s$-model action (\NLSa) becomes the derivative-free expression
$$ {\cal L}_\s~=~ {2\over\a^2} b_\mu b^{*\mu}\eq $$
By substituting (\rotU) into (\topoC) we find that
the topological current is again given by (\auxGF) in terms of
$a_\mu$. This allows us to identify $a_\mu$ with the auxiliary gauge field of
the $CP^1$ formulation.

The global $su(2)$ symmetry of the $\s$-model is naturally embedded in this
formulation since the field $B_\mu$ is invariant under an $su(2)$ rotation
$\p\to V^{-1}\p V$. On the other hand, the arbitrariness of $U$ gives rise
to a local Abelian gauge symmetry for which $a_\mu$ is the gauge field:
$$\eqalign{
a_\mu~&\to~ a_\mu+{1\over 2\pi}\d_\mu\b(x)\cr
b_\mu~&\to~ e^{i\b(x)}b_\mu \cr
\chi~&\to~ e^{-i\b(x)\s_3}\chi \cr} \eq $$

The auxiliary, curl-free gauge field $B^a_\mu$ couples minimally to the rotated
fermion $\chi$. The change of variables from $\psi$ to $\chi$ does not give
rise to any nontrivial Jacobian since $U$ is unitary.
Finally, we may incorporate the electromagnetic field into $B_\mu$ by defining
$B^0_\mu = e_0A_\mu$ and $\s_0=1$.

\subsec{Parity Transformations}

Some remarks are in order concerning parity. In $(2+1)$ dimensions the parity
transformation is defined as the reversal of one of the spatial coordinates,
say $x$.
Its effect on the fermion $\chi$ is defined by a matrix $P$:
$$ \chi(x,y,t)\rightarrow P\chi(-x,y,t) \eq $$
chosen so as to keep the fermion kinetic term invariant. Consequently $P$ must
obey the following constraints:
$$ \eqalign{
P^\dg\g^0\g^1 P &= -\g^0\g^1\cr
P^\dg\g^0\g^2 P &= \g^0\g^2\cr
P^\dg P &= 1\cr}\eq $$
The solution to these constraints is $P=\Pi\g^1$, where $\Pi$ is a unitary
matrix acting only on spin space.
The $su(2)$ invariant fermion mass term $m\bar\psi\psi$ becomes
$m\bar\chi\chi$ in terms of the rotated fields.
Such a mass term violates parity. Indeed, a fermion mass term
$\bar\chi M\chi$ conserves parity if $\Pi^\dg M\Pi=-M$, which admits the
solution $M= m\s_3$ (opposite masses for the two spin components, given
here by the `rotated' Yukawa coupling) and $\Pi=\s_1$ (or $\s_2$).
{}From this we easily work out the parity transformations of the vector fields:
$$\eqalign{A_{0,2}&\to A_{0,2}\cr a_{0,2}&\to -a_{0,2}\cr
b_{0,2}&\to b^*_{0,2}\cr}\qquad\quad
\eqalign{A_1&\to -A_1\cr a_1&\to a_1\cr b_1&\to -b^*_1\cr} \eq $$
The Lagrangian (\FerB) then manifestly conserves parity.
Notice that $a_\mu$ and $A_\mu$ have opposite intrinsic parities.

\subsec{Radiative Corrections}

The one-loop effective action for the gauge field $B_\mu$ (which now
includes $A_\mu$ as one of its components) may be obtained from a
single Feynman diagram, thanks to gauge invariance.
The vacuum polarization $\Pi^{\mu\nu}$ has the form
$$\eqalign{
\Pi^{\mu\nu}_{ab} &= -\int{d^3k\over(2\pi)^3}\tr\l\{
\s_a\g^\mu{\Sl{p}+\Sl{k}-g\s_3\over (p+k)^2-g^2}\s_b\g^\nu
{\Sl{k}-g\s_3\over k^2-g^2}\r\}\cr
&= -i\Pi_{ab}(p^2)(g^{\mu\nu}p^2-p^\mu p^\nu) ~+~
\Pit_{ab}(p^2) \e^{\mu\nu\la}p_\la\cr} \eq $$
where the lower indices, running from 0 to 3, pertain to the internal
components of $B_\mu=B_\mu^a\s_a$.
Following the usual methods of integration, and in particular using the
identity $\tr\g^\mu\g^\nu\g^\la=-2i\e^{\mu\nu\la}$, we arrive at
$$ \Pi_{ab}(0) = {1\over 6\pi|g|}\delta_{ab} \qquad
\Pit_{03}(0) = -{1\over 2\pi}\sign(g) \eq $$
Other components vanish.
The above vacuum polarization forces a charge renormalization $e_0\to e$
and a rescaling $e_0 A_\mu\to eA_\mu$ of the electromagnetic field.
The effective Lagrangian, in terms of $A_\mu$ and $a_\mu$, is then
$$\eqalign{{\cal L}_{eff}&=
-\frac14\l({1\over 12\pi\sqrt{\lm_0}}+(2\pi/e)^2\r)f_{\mu\nu}f^{\mu\nu}\cr
&\qquad -\frac14 F_{\mu\nu}F^{\mu\nu}
-e~\sign(g)\e^{\mu\nu\la}A_\mu\d_\nu a_\la \cr} \eq $$
We see that the induced electromagnetic charge of the $\s$-model solitons
is $q = e~\sign(g)$.
Let us assume for simplicity that $|g|\ll e_0^2,\sqrt{\lm_0}$.
Then the fermion-induced polarization dominates and we must apply the following
renormalization and rescaling in order to recover the form (\CSb):
$$\eqalign{
e_0 ~\to~ & e=\sign(e_0)\sqrt{6\pi|g|}\cr
a_\mu~\to~ &{e\over 2\pi}a_\mu\cr}\eq $$
Accordingly, the Chern-Simons mass is $m=e^2/2\pi$.
Note that the Yukawa coupling constant $g$ has disappeared from our result.
It could in fact be as small as we wish; the final result would be the same in
terms of the renormalized charge $e$.

\newsec{Discussion and Stability of the Result}

The essential feature of the above calculation is the appearance of
electrically charged solitons at one-loop.
The induced current-current term $j^\mu j_\mu$ is expected to be generic,
i.e. its existence should not depend on the type of fermion interaction
used, since it is already present within the $\s$-model alone.
On the other hand, it is not a priori clear that the induced soliton charge
exists for all kinds of fermion interactions.

In order to shed some light on this question, let us calculate the induced
soliton charge again, but this time within a more general formalism.
Since the calculation is already published \HST, the derivation will be
simply highlighted.
Let us write the tree-level fermion action as
$$ \chi^\dagger S^{-1}(p-B)\chi \eq $$
where $S^{-1}(p)$ is the inverse fermion propagator as a function of momentum.
In this notation $\chi$ is a vector in configuration space, and $S$ is a
matrix. A functional trace (i.e. a momentum integration with a trace over
Dirac and spin indices) is understood. The 1-loop effective action is then
$$\eqalign{
S_{eff} ~&=~ -i~\Tr~\log~S^{-1}(p-B)\cr
&=~ i\sum_{n=1}^\infty {1\over n}~\Tr\l[ S(p)V(p,B)\r]^n \cr}\eq $$
where $V(p,B)=S^{-1}(p)-S^{-1}(p-B)$ is the interaction vertex with the
external field $B_\mu(x)$.

This series contains an infinite number of terms, with an arbitrarily high
number of derivatives. In the long wavelength approximation, only the lowest
terms are relevant. In particular, we wish to extract the coefficient $q$ of
$$ \int d^3x~\e^{\mu\nu\la}A_\mu\d_\nu a_\la ~=~
{1\over 2\pi e}\int d^3x~ \e^{\mu\nu\la}B^0_\mu\d_\nu B^3_\la \eq $$
This can be done by a derivative expansion. Details of the calculation can be
found in \HST; here we simply state the result:
$$ {q\over e}~=~{\pi\over3}\e^{\mu\nu\la}\int{d^3p\over (2\pi)^3}~\tr\l\{
\s_3 S{\d S^{-1}\over\d p^\mu} S{\d S^{-1}\over\d p^\nu}
S{\d S^{-1}\over\d p^\la} \r\}\eq $$
This expression reminds us of a Wess-Zumino term, except for the
fact that it is integrated in momentum space, and that $S$ is a Hermitian (as
opposed to unitary) matrix. We may immediately conclude that $q$ vanishes if
$S$ is completely diagonal, and in particular if it lacks a Dirac matrix
structure. Let us write $S^{-1}$ as
$$ S^{-1}(p) ~=~p_0 + f(p_i) - \omv(p_i)\cdot\vec\tau \eq $$
where $f$ and $\omv$ are arbitrary functions of spatial momentum, and where
$\tau^a$ are the Pauli matrices acting on Dirac space. For the specific model
(\FerB), $f=0$ and $\omv=(p_2,-p_1,g\s_3)$. We assume that $S^{-1}$ is
diagonal in spin space, i.e., involves only $\s_3$.
The unit vector function
$\omh=\omv/|\omv|$ maps the momentum plane onto the two-sphere. Let
$\Omega_r$ $(r=1,2)$ be the fraction of total solid angle covered by this
mapping for the spin component $r$. Then it is simple to show \HST\ that
$$ q/e = \Omega_1-\Omega_2 \eq $$
These solid angles are in general determined by the asymptotic behavior of
the fermion dispersion relation as $\vec{p}\to 0$ and $\vec{p}\to\infty$.
In the case of (\FerB),
$$ \omh(0) ~=~(0,0,\s_3~\sign(g))\quad\qquad \omh(\infty)
{}~=~(\cos\theta,-\sin\theta,0) \eq $$
where $\theta$ is the azimuthal angle on the momentum plane.
It follows that $\Omega_1=\hf\sign(g)$, $\Omega_2=-\hf\sign(g)$,
and the induced soliton charge is
$$  q = e~\sign(g) \eq $$

This specific result may also be obtained by tracking the number of energy
levels crossing the Fermi sea as a soliton background is adiabatically turned
on
\CCW. The method sketched above has the advantage of clearly demonstrating the
stability of the induced charge with respect to perturbations of the fermion
dispersion relation and the inclusion of more interactions.
Indeed, the coupling between the rotated fermion $\chi$
and $a_\mu$ or $A_\mu$ will always be minimal, with the correct coefficient,
irrespective of other couplings that we might add between $\vph$ and $\psi$,
such as $\bar\psi[\p,\Sl\d\p]\psi$, $\bar{\psi}\Sl\d\p\psi$ and so on.
This method shows, however, that the existence of a Dirac matrix structure is
essential.
Finally, let us mention that the induced soliton charge, being naturally
quantized, does not receive higher-loop corrections, in accordance with
the theorem of Coleman and Hill \CH.
\medskip
\centerline{\bf Acknowledgments}
\medskip
This work is supported by NSERC (Canada) and F.C.A.R. (Qu\'ebec).

\listrefs
\bye